\documentclass[conference]{IEEEtran}

\usepackage{cite}
\usepackage{amsmath,amssymb,amsfonts}
\usepackage{algorithmic}
\usepackage{graphicx}
\usepackage[dvipsnames]{xcolor}
\usepackage[final]{microtype}
\usepackage[italic]{mathastext}
\usepackage{libertine}
\usepackage[T1]{fontenc}
\usepackage{textcomp}
\usepackage[varqu,varl]{zi4}
\usepackage[all]{nowidow}
\usepackage[keeplastbox]{flushend}
\usepackage{fancyhdr}

\usepackage{color}

\usepackage{microtype}
\usepackage{xspace}
\usepackage{graphicx}
\usepackage{subfigure}
\usepackage{booktabs} 

\usepackage{multirow}
\usepackage{comment}

\usepackage{hyperref}
\usepackage{bm}

\usepackage{todonotes}

\usepackage{pifont}

\usepackage{amsmath,amsfonts}
\usepackage{algorithmic}
\usepackage{graphicx}
\usepackage{textcomp}
\usepackage{xcolor}
\usepackage{multirow}
\RequirePackage[nameinlink]{cleveref} 
\crefname{chapter}{Chapter}{Chapters}
\crefname{section}{Section}{Sections}
\crefname{subsection}{Section}{Sections}
\crefname{equation}{Equation}{Equations}
\crefname{definition}{Definition}{Definitions}
\crefname{assumption}{Assumption}{Assumptions}
\crefname{theorem}{Theorem}{Theorems}
\crefname{figure}{Figure}{Figures}
\crefname{table}{Table}{Tables}
\crefname{algorithm}{Algorithm}{Algorithms}
\let\autoref\cref 

\usepackage{multirow} 
\usepackage{multicol} 
\usepackage{makecell} 

\usepackage{booktabs}   
\usepackage{tabularx}   
\usepackage{hyperref}   
\usepackage{graphicx}   
\usepackage{longtable}
\usepackage{adjustbox}
\usepackage{longtable}

\usepackage{enumitem}

\usepackage{authblk}

\newcommand{\insertFigure}[2]{
    \begin{figure}[t]
        \centering
        \includegraphics[width=\linewidth]{\FIGDIR/#1.pdf}
        \vspace{-7mm}
        \caption{#2}
        \vspace{-4mm}
        \label{fig:#1}
    \end{figure}
}

\newcommand{\insertWideFigure}[2]{
    \begin{figure*}[t]
        \centering
        \includegraphics[width=\textwidth]{\FIGDIR/#1.pdf}
        \vspace{-6mm}
        \caption{#2}
        \vspace{-4mm}
        \label{fig:#1}
    \end{figure*}
}

\ifx\forsubmission\undefined
\newcommand{\TODO}[1]{\textcolor{red}{TODO: #1}}
\newcommand{\SK}[1]{\textcolor{red}{SK: #1}}
\newcommand{\HK}[1]{\textcolor{blue}{HK: #1}}
\newcommand{\RK}[1]{\textcolor{brown}{RK: #1}}

\newcommand{\fixme}[1]{{\color{red} {#1}}}

\else
\newcommand{\TODO}[1]{\textcolor{red}{}}
\newcommand{\SK}[1]{\textcolor{red}{}}
\newcommand{\FK}[1]{\textcolor{blue}{}}
\newcommand{\HK}[1]{\textcolor{blue}{}}
\newcommand{\RK}[1]{\textcolor{green}{}}

\newcommand{\fixme}[1]{{\color{red} {}}} 

\fi

\newcommand{\squishlist}{
 \begin{list}{$\bullet$}
  { \setlength{\itemsep}{0pt}
     \setlength{\parsep}{3pt}
     \setlength{\topsep}{3pt}
     \setlength{\partopsep}{0pt}
     \setlength{\leftmargin}{1.5em}
     \setlength{\labelwidth}{1em}
     \setlength{\labelsep}{0.5em} } }

\newcommand{\squishlisttwo}{
 \begin{list}{$\bullet$}
  { \setlength{\itemsep}{0pt}
     \setlength{\parsep}{0pt}
    \setlength{\topsep}{0pt}
    \setlength{\partopsep}{0pt}
    \setlength{\leftmargin}{2em}
    \setlength{\labelwidth}{1.5em}
    \setlength{\labelsep}{0.5em} } }

\newcommand{\squishend}{
  \end{list}  }

\newcommand{\betterparagraph}[1]{\noindent \textbf{#1. }}

\usepackage{placeins}
\usepackage{cite}
\usepackage{amsmath,amssymb,amsfonts}
\usepackage{algorithmic}
\usepackage{graphicx}
\usepackage{textcomp}
\usepackage{xcolor}

\def\FIGDIR{./Figures}

\def\BibTeX{{\rm B\kern-.05em{\sc i\kern-.025em b}\kern-.08em
    T\kern-.1667em\lower.7ex\hbox{E}\kern-.125emX}}

\begin{document}



\title{Characterizing State Space Model and Hybrid Language Model Performance with Long Context}

\author{Saptarshi Mitra \ \ Rachid Karami \ \ Haocheng Xu \ \ Sitao Huang
\ \ Hyoukjun Kwon}
\affil{
\textit{Electrical Engineering and Computer Science}\\
\textit{University of California, Irvine}\\
Irvine, USA\\
saptarshi.mitra@uci.edu}


\maketitle

\begin{abstract}

Emerging applications such as AR are driving demands for machine intelligence capable of processing continuous and/or long-context inputs on local devices.
However, currently dominant models based on Transformer architecture suffers from the quadratic computational and memory overhead, which hinders applications required to process long contexts.
%
This has spurred a paradigm shift towards new architectures like State Space Models (SSMs) and SSM-Transformer hybrid models, which provide near-linear scaling.
The near-linear scaling enabled efficient handling of millions of tokens while delivering high performance in recent studies.
Although such works present promising results, their workload characteristics in terms of computational performance and hardware resource requirements are not yet thoroughly explored, which limits our understanding of their implications to the system level optimizations.


%

To address this gap, we present a comprehensive, comparative benchmarking of carefully selected Transformers, SSMs, and hybrid models specifically for long-context inference on consumer and embedded GPUs.
%
%
Our analysis shows that SSMs are well-suited for on-device AI on consumer and embedded GPUs for long context inferences.
While Transformers are up to 1.9$\times$ faster at short sequences (<8K tokens), SSMs demonstrate a dramatic performance inversion, becoming up to 4$\times$ faster at very long contexts (\textasciitilde57K tokens), thanks to their linear computational complexity and \textasciitilde64\% reduced memory footprint.
%
Our operator-level analysis reveals that custom SSM kernels like selective scan despite being hardware-aware to minimize memory IO, dominate the inference runtime on edge platforms, accounting for over 55\% of latency due to their sequential, element-wise nature.
%
%
%
%
To foster further research, we are sharing CPU/GPU profiling traces and have made our characterization framework SSM-Scope open-sourced at https://github.com/sapmitra/ssm-scope
%
\end{abstract}
\section{Introduction}
\label{sec:intro}

Recent advancements in large language models (LLMs)~\cite{Touvron2023LLaMAOA, achiam2023gpt} have enabled strong general problem solving capabilities to artificial intelligence (AI) models spanning both discriminative (e.g., multiple-choice question answering~\cite{hendrycks2021measuring}) and generative (e.g., code generation\cite{liu2023your}) tasks.
%
%
The context length, which is the length of the input text to an LLM, is trending to increase to reduce hallucinations and maintain output coherence~\cite{tang2025seeing}.
Also, long context is crucial for emerging applications such as long document processing and retrieval tasks~\cite{chenlong2023lora}.
%
However, Transformer-based LLMs have quadratic computational and memory complexity on context length, limiting their ability to handle long contexts.

To address the challenge, new models with selective state mechanisms, or state space models (SSMs), has been proposed~\cite{gu2024mambalineartimesequencemodeling}.
SSMs~\cite{gu2024mambalineartimesequencemodeling} and Transformer-SSM hybrid models~\cite{nvidiaNemotronHFamilyAccurate2025, behrouz2024titans} enable a linear scaling in the context length, which dramatically reduces both the computational cost and the memory footprint by eliminating the need for a large Key-Value (KV) cache.
%
Such architectural innovation is breaking through previous computational complexity and memory barriers, enabling models with long context windows on consumer and even embedded hardware.
To motivate our study, \autoref{fig:intro_ttft_4} encapsulates the performance trade-off between these architectures by showing the Time-to-First-Token (TTFT) and Time-per-Output-Token (TPOT) for two similarly sized models on a consumer GPU (RTX 4090).
At shorter sequence lengths, the highly optimized Transformer-based model, Qwen2.5, outperforms the State-Space-Model, Mamba2, by a factor of 1.9$\times$ and 1.1$\times$, respectively, for both TTFT and TPOT, as shown in \autoref{fig:intro_ttft_4} (a) and (c). 
However, as the input context is scaled to a higher length of 32K tokens, this performance dynamic inverts, with Mamba2 outperforming the Transformer by 2.65$\times$ and 3$\times$ respectively for TTFT and TPOT, as shown in \autoref{fig:intro_ttft_4} (b) and (d).
This result highlights a fundamental scaling advantage of SSMs; beyond this context length, Mamba2 can continue to operate with manageable latency, whereas the Transformer model cannot, a limitation we analyze in detail in \autoref{subsec:scalability_mem_footprint}.

\insertFigure{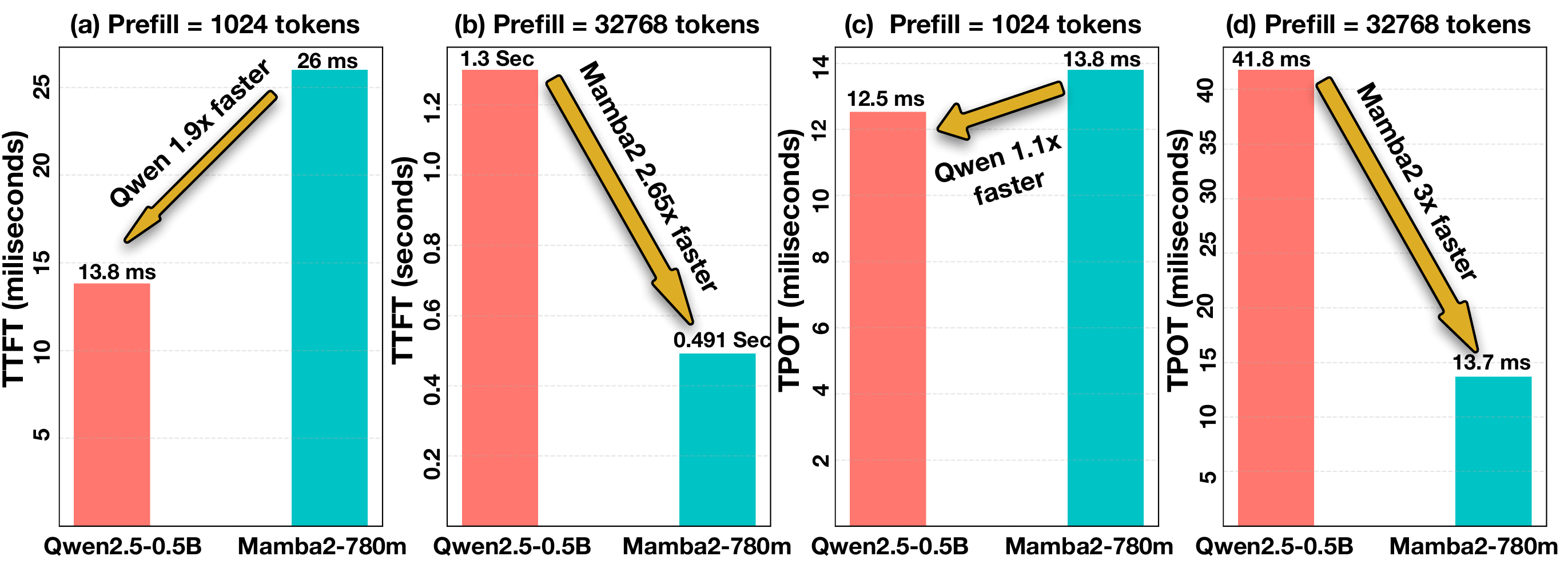}{TTFT (a,b) and TPOT (c,d) scaling comparison of Qwen2.5-0.5B \cite{yang2025qwen3} and Mamba2-780m \cite{dao2024transformers}.  While Qwen is faster (1.9$\times$) at shorter sequence lengths, Mamba2's superior scaling provides a significant performance advantage(2.65$\times$) at longer contexts for both prefill and decode (for generation length 256 with batch size 1) stages.}

Because of the strong scaling over the context length, the SSM and hybrid models are being adopted in the industry.
%
%
%
For instance, the recently released Falcon-H1\cite{zuo2025falcon} series uses a parallel hybrid architecture combining Transformer attention with Mamba-2 SSMs, enabling models as small as 0.5B parameters to match the performance of 7B baselines from 2024 while supporting 256K context windows.
Similarly, NVIDIA's Hymba\cite{dongHymbaHybridheadArchitecture2024} introduced a novel hybrid-head design that integrates attention and SSM heads within the same layer, achieving over 11$\times$ reduction in KV-cache size compared to a similar-sized Transformers.
%
More recently, Nemotron-Flash\cite{fu2025nemotron} optimized these hybrid primitives for real-world edge latency, delivering over 45$\times$ higher throughput than comparable Transformer models by strategically combining attention, Mamba-2, and DeltaNet operators.

Although these new models show immense promise, a deep, quantitative understanding of their performance characteristics, operational bottlenecks, and scaling behavior on consumer devices is critically lacking.
%
%
To address this gap and guide future research in hardware and systems design for edge AI, we conduct a comprehensive performance characterization of state-of-the-art SSM and hybrid models.
%
%
%
We aim to highlight the opportunities and challenges of deploying the next generation of long-context LLMs, focusing on workstation and edge device deployment scenarios facilitating local inference for privacy and latency.
The key contributions of this work are as follows:
\begin{itemize}[leftmargin=*, labelsep=5pt]
{\item We present comprehensive memory footprint analysis of Transformer, SSM, and hybrid models for extremely long-context inference (up to 220K tokens) on consumer and embedded GPUs.
We establish the out-of-memory (OOM) frontier, demonstrating that optimized Transformers hit memory limits at \textasciitilde65K tokens, while SSMs operate effectively at 4$\times$ longer sequences without offloading on the consumer GPU.}
{\item We quantify the performance and energy trends with respect to increasing sequence lengths between Transformers and SSMs. At 57K tokens, the quadratic complexity of Transformers results in prohibitive energy consumption (1492 J), whereas an SSM reduces this by \textasciitilde75\% (370 J).}
{\item We conduct a detailed, operator-level, cross-device latency characterization of these models, identifying the performance bottlenecks (SSM-kernels dominate >55\% of total latency on edge platforms) and distinct runtime profiles of novel operators.}
\end{itemize}

\section{Background}
\label{sec:background}

\insertFigure{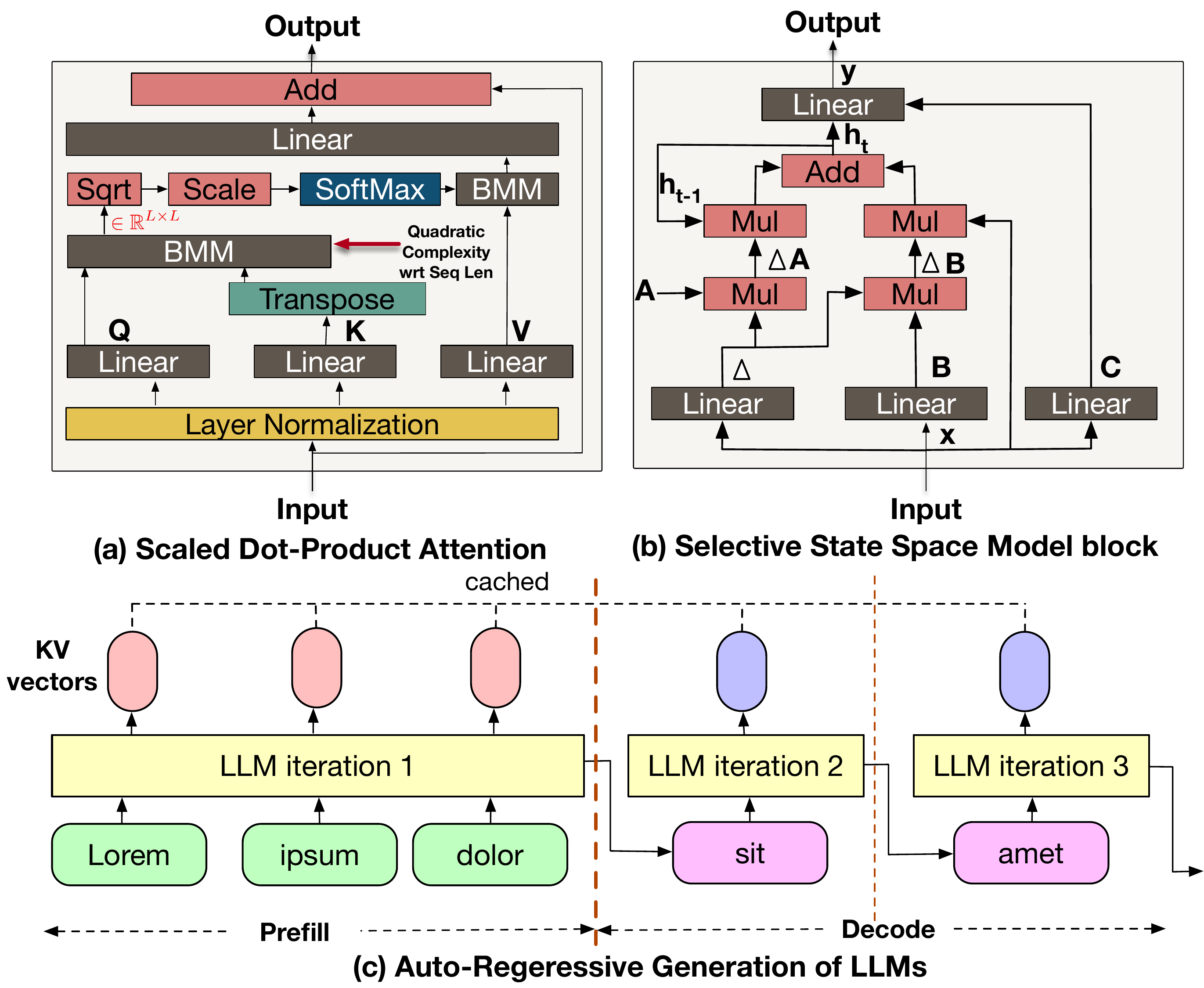}{(a) Basic building block of Transformers: a scaled dot-product attention module; (b) a S6 block showing fundamental computation of SSMs. (c) Overview  of auto-regressive generation (prefill \& decode) of LLMs}


\subsection{Language Model Architectures}
\label{subsec:lm_architecture}

We discuss three model architecture classes for language models: Transformer, SSM, and Hybrid model.

\betterparagraph{Transformers} The landscape of large language models (LLMs) has been predominantly shaped by the Transformer architecture~\cite{vaswani2023attentionneed}.
Its success stems from the self-attention mechanism, enabling effective modeling of long-range dependencies and demonstrating remarkable scaling properties~\cite{DBLP:journals/corr/abs-2001-08361}.
%
%
However, during inference, the Transformer is bottlenecked by the ($O(N^2)$) complexity of self-attention, where $N$ is the sequence length.  
%
Transformer-based LLMs rely on the KV cache\cite{pope2022efficientlyscalingtransformerinference} for inference acceleration to store and reuse representations of past tokens, avoiding redundant computations.
However, the memory footprint of this cache scales linearly ($O(N)$) with sequence length $N$, necessitating costly offloading to slower memory tiers, introducing significant latency bottlenecks that hinder real-time generation and deployment on resource-constrained devices. 
%
%
\autoref{fig:background}(a) shows a scaled dot-product attention module with all the computations involved.
During inference, LLMs operate in two distinct phases.
The initial \textit{prefill} phase processes the entire input prompt simultaneously, making it a highly parallel task involving extensive matrix computations that make the system compute-bound.
This is followed by the \textit{auto-regressive decode} phase, which generates the output one token at a time, iteratively using the previously generated token as input for the next step, as shown in \autoref{fig:background}(c).
%
The decode phase is typically memory-bound because each token generation relies on all previous tokens, and transferring previously computed Keys (K) and Values (V) from the KV cache dominates as the sequence length grows.

\betterparagraph{State Space Models (SSMs)}\label{SSM_ops}To overcome efficiency and scalability challenges of Transformers with long sequences,
SSMs~\cite{gu2022efficientlymodelinglongsequences}, especially recent variants like Mamba ~\cite{gu2024mambalineartimesequencemodeling} and Mamba-2 ~\cite{dao2024transformers}, have emerged as promising alternatives.
%
%
SSMs offer linear time complexity ($O(N)$) in sequence length and constant memory ($O(1)$) during autoregressive generation via a recurrent state representation.
They can be efficiently parallelized during training, similar to Transformers, but operate like RNNs during inference.
A simplified Selective State Space Model (S6) block\cite{gu2024mambalineartimesequencemodeling}, as shown in \autoref{fig:background}(b) computes an output sequence $y$ from an input sequence $x$ by modeling a latent state $h$. 
Here, $\textbf{A}$ is derived from a fixed continuous-time parameter, while the selective parameters $\textbf{B}$ and $\textbf{C}$ are dynamically generated from the input sequence at each step.
\(\Delta\) represent the decay rate, which defines how the influence of the current state ($h_t$) diminishes with time.

%
When discretized, the core computation can be expressed in a concise form by the following state equations (k: current time step):
\begin{equation}
\begin{aligned}
    h_{k+1} &= \bar{A} h_k + \bar{B}_k x_k \\
    y_k &= \bar{C}_k h_k
\end{aligned}
\label{eq3}
\end{equation}
%
%
%
%
%
%
%
%
%
The inherent recurrent dependencies of SSM limit Instruction Level Parallelism (ILP).
Parallel prefix-scan\cite{harris2007parallel} algorithms require heavy thread synchronization, reducing warp occupancy compared to GEMMs.
This results in a lower Arithmetic Intensity (FLOPs/B) than dense GEMMs, making the core SSM kernels memory-bound rather than compute-bound.
Current SSM-scan implementations in Mamba2\cite{dao2024transformers} (SSD) depend heavily on kernel-fusion to overcome this memory-bound nature. 
However, fused SSM kernels generate massive intermediate data (e.g., 642KB vs 321KB for FlashAttention-2), exceeding the Streaming Multiprocessor (SM) shared memory capacity (128KB and 164KB for RTX4090 and A100 respectively) and forcing spills to global memory \cite{jung2025hlx}.
This leads to high register pressure and shared memory contention.
Additionally, the SSM-scan is dominated by element-wise Hadamard products and frequent memory movements.
Despite these microarchitectural bottlenecks, we present a performance inversion for long-contexts where the Transformer's quadratic memory traffic eventually outweighs the SSM's inefficient use of on-chip compute resources.

While demonstrating strong performance on various accuracy benchmarks, pure SSM architectures have shown limitations, particularly in tasks requiring precise information retrieval or complex in-context learning (ICL) compared to attention-based models~\cite{10.5555/3692070.3693682, waleffeEmpiricalStudyMambabased2024}.
%
Our evaluation in \autoref{fig:accuracy_ttft_2} shows a significant accuracy gap in in-context reasoning tasks like 5-shot MMLU, where Mamba-2 lags behind Qwen2.5 by about 25 points (36.3\% vs. 61.1\%).

\insertFigure{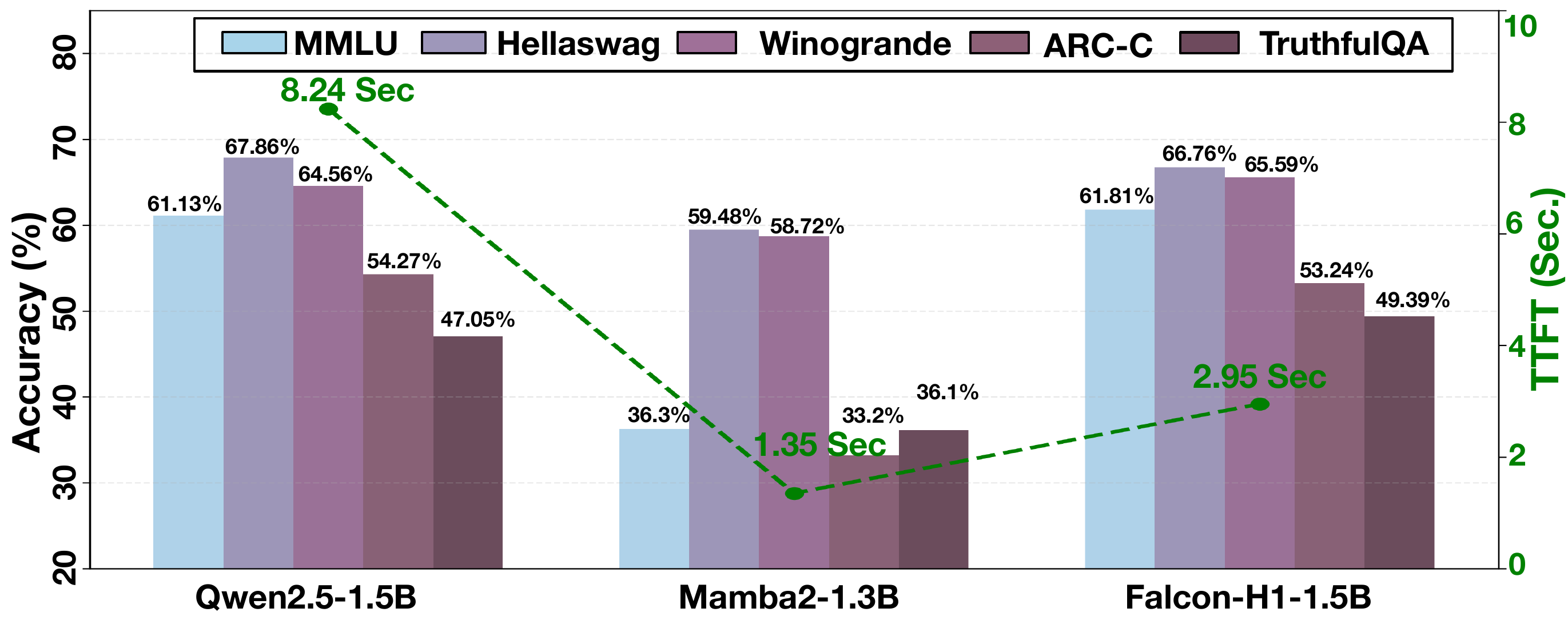}{Accuracy-latency efficiency frontier analysis of Transformer (Qwen2.5), SSM (Mamba2), and Hybrid (Falcon-H1) models of similar size ($\approx$1.5B) for 57K sequence length during prefill stage.}

\betterparagraph{Hybrid Architectures} 
Recent research has focused on developing hybrid architectures that integrate the complementary strengths and weaknesses of Transformers and SSMs.
Hybrid architectures can be categorized by inter-layer (sequential) or intra-layer (parallel) fusion.
Sequential hybrids, like Zamba2\cite{gloriosoZamba2SuiteTechnical}, alternate softmax attention and SSM layers at specific intervals. Parallel hybrids, like Hymba\cite{dongHymbaHybridheadArchitecture2024} and Falcon-H1\cite{zuo2025falcon}, blend attention and SSM heads in parallel within individual layers for fine-grained fusion.
The goal is to retain the efficiency and sequence modeling capabilities of SSMs while leveraging the proven strengths of attention for tasks requiring precise recall and complex reasoning over context.
%
To motivate the shift towards hybrid architectures, we benchmarked representative models from each category: Transformer (Qwen2.5-1.5B), SSM (Mamba2-1.3B), and Hybrid (Falcon-H1-1.5B) across diverse tasks such as MMLU (knowledge \& reasoning)\cite{hendrycks2021measuring}, Hellaswag (commonsense reasoning)\cite{zellers-etal-2019-hellaswag}, Winogrande (Ambiguity Resolution)\cite{10.1145/3474381}, ARC-C (Deep reasoning)\cite{clark2018think} and TruthfulQA (long-context truthfulness)\cite{lin-etal-2022-truthfulqa}.
Our evaluation reveals that the SSM offers the fastest inference with a TTFT of 1.35s for a 57K prefill, but suffers a significant performance penalty on knowledge-intensive benchmarks. 
For instance, on TruthfulQA, Mamba2 achieves only 36.1\% accuracy, lagging well behind the Qwen2.5 Transformer (47.05\%).
The hybrid Falcon-H1 model, however, not only bridges this gap but surpasses the Transformer baseline with an accuracy of 49.39\%.
Similar trend is also observed in Winogrande task where the hybrid model (65.59\%) effectively matches the Qwen model (64.56\%) performance.
Crucially, the hybrid architecture achieves this accuracy parity while maintaining a 2.8$\times$ speedup in TTFT (2.95s) over the Transformer (8.24s), making it a Pareto-optimal solution for efficient, on-device inference.
%


\subsection{Memory Requirements for LLM inference}
%
The usability of a large language model on any deployment platform is often constrained by its memory usage during inference.
%
%
The primary contributors to this memory footprint, especially for Transformer models, apart from the model weights ($N_{\text{params}}\times p$; $p$ is Bytes per element) are KV cache and activation memory, whose memory footprints are calculated as follows (\textbf{\textit{B}}: batch size, \textbf{\textit{S}}: sequence length, \textbf{\textit{L}}: Number of layers, \textbf{\textit{D}}: hidden\ dimension, \textbf{\textit{C}}: Number of layers to keep their activations on memory):
    \begin{equation}
        \text{Memory}_{\text{KV-cache}} = B \times S \times L \times D \times 2 \times 
        p
    \end{equation}

    \begin{equation}
    \label{eq:activation_memory}
        \text{Memory}_{\text{activations}} \approx B \times S \times D \times C \times 
        p
    \end{equation}

Note that the KV-cache memory footprint can vary depending on optimizations (e.g., grouped query attention; GQA~\cite{ainslie-etal-2023-gqa}).
Here we focused on off-the-shelf models as publicly available representative models: Qwen2.5 family that includes GQA.

The total memory footprint is the sum of the three components (weight, KV cache, and activation) and the additional overhead reserved by the inference framework.
The choice of framework significantly impacts memory management and OOM thresholds. 
Advanced engines like vLLM use sophisticated techniques like PagedAttention\cite{10.1145/3600006.3613165} to mitigate memory fragmentation, while the basic HuggingFace Transformers pipeline used in our experiments may exhibit different memory characteristics.
Memory offloading techniques\cite{10.1145/3721146.3721961}\cite{10.5555/3618408.3619696}\cite{lee2024infinigen} that strategically shifts data within different memory hierarchies were also \textit{not} considered to identify inherent properties of the models in simplest runtime environment.
\subsection{Computations in Various Language Models: ML Operators}
Modern machine learning models are composed of a wide array of computational operators, with their types and distributions varying significantly across architectures.
The operators can be broadly categorized into three distinct families:
\textit{(a) GEMM-based Operators}: GEMM-based operators refer to operators mainly driven by General Matrix Multiplication (GEMM), such as  \texttt{Conv2D} and \texttt{Linear}, which have traditionally used as building blocks to construct backbones of ML models.
%
%
Due to their dominance in the operator counts and computational intensity, they often dominate the execution time.
However, thanks to their regular computational graph, various hardware and software optimizations have been explored for acceleration. 
Extensive research has shown that specialized hardware accelerators \cite{Thesis_Saptarshi_Mitra_2a_print.pdf} or efficient runtime kernels \cite{Ganji_2023_CVPR} can significantly reduce the latency of GEMM-centric operations on resource-constrained edge devices.
\textit{(b) Non-GEMM Operators}: Non-GEMM operators refer to all the operators that is not primarily GEMM-based, such as memory layout manipulations (e.g., \texttt{reshape}, \texttt{transpose}), normalization functions, and activations.
As GEMM performance has been heavily optimized, non-GEMM operators have emerged as a new performance bottleneck\cite{karami2025nongemmbenchunderstandingperformance}.
%
%
The specific non-GEMM operators that dominate runtime vary considerably depending on the model and task, highlighting their heterogeneity.
\textit{(c) SSM-Specific Operators}: Newer architectures introduce another class of unique, specialized operators.
These kernels, implementing the core selective scan mechanism, have distinct computational properties that differ from traditional GEMM or non-GEMM ops, which will be discussed in \autoref{subsec:op_breakdown}.

To understand the landscape of the memory requirement and computational performance of recent language models (Transformer, SSM, and Hybrid models) that contain the ML operators, we build a performance charcterization flow, which we discuss next.

\section{Methodology}
\label{sec:methodology}


%
To ensure fair and comprehensive benchmarking and profiling, we considered (a) similarly sized language models, recently released SSMs and hybrids that accurately represent the current state of long-sequence data handling, (b) practical deployment scenario with off-the-shelf consumer compute devices like workstation and edge GPUs, (c) precisely capturing operator-level performance metrics to represent the usability of different models in varied applications and (d) utilizing a wide range of workload configurations to generate meaningful insights in real-world deployment.
We carefully curated our model suite for fair evaluation and cross-platform compatibility in BF16.
All models were sourced directly from official HuggingFace repositories or their author-provided GitHub implementations to maintain consistency and reproducibility.
%
%
The hardware platforms and their respective setups are described in the following \autoref{subsec:platforms}.

\begin{table}[t]
\centering
\caption{Evaluation Setup for Consumer and Embedded GPU Platforms.}
\label{tab:gpu_setup}
\begin{tabular}{@{}lcc@{}}
\toprule
\textbf{Specification} & \textbf{GPU 1 (Consumer)} & \textbf{GPU 2 (Edge)} \\ \midrule
GPU Model                        & NVIDIA RTX 4090           & NVIDIA Jetson Orin Nano$^1$   \\
Architecture                     & Ada Lovelace              & Ampere                    \\
SMs  & 128                       & 8                         \\
Comp. Throughput           & $\sim$330 TFLOPS          & $\sim$20 TFLOPS           \\
GPU Memory                       & 24 GB GDDR6X              & 8 GB LPDDR5 (Shared)      \\
Memory Bandwidth                 & 1008 GB/s                 & 68 GB/s                   \\
Host Interconnect                & PCIe 4.0, 16x             & Integrated on-chip        \\ \bottomrule
\multicolumn{3}{l}{\footnotesize $^1$ \texttt{MAXN} power mode enabled with 16GB swap from NVMe}
\end{tabular}
\end{table}
\subsection{Evaluation Platforms}
\label{subsec:platforms}
%
%
%
In the current technological landscape, the high performance of leading SSMs is primarily achieved through hardware-aware algorithms implemented as custom CUDA kernels, making them highly efficient on modern GPUs.
%
%
Our evaluation is conducted on two distinct GPU platforms, representing a consumer desktop and a power-constrained edge system.
%
%
The detailed specifications of our evaluation platforms are presented in ~\autoref{tab:gpu_setup}.
We primarily use PyTorch~\cite{paszke2019pytorch}, along with CUDA Toolkit 12.x (depending on model-specific kernel availability), for all experiments.

\begin{table}[h!]
\centering
\caption{The suite of models selected for characterization.\\
\textbf{Quant.:} Availability of official INT4/8 checkpoints.}
\label{tab:model_zoo_short}
\resizebox{\columnwidth}{!}{
\begin{tabular}{@{}llll@{}}
\toprule
\textbf{Arch.} & \textbf{Family} & \textbf{Parameters (Sizes)} & \textbf{Quant.} \\ \midrule
\multirow{5}{*}{Trans.} & Qwen2.5 & 0.5B, 1.5B & \checkmark \\
 & Phi-3-mini & 3.82B & \checkmark \\
 & Llama-3.2 & 1B & \checkmark \\
 & TinyLlama & 1.1B & \checkmark \\
 & GPT-neo & 125M & -- \\ \midrule
\multirow{2}{*}{SSM} & Mamba & 130M, 370M, 790M, 1.4B, 2.8B & \checkmark \\
 & Mamba-2 & 130M, 370M, 780M, 1.4B, 2.8B & \checkmark \\ \midrule
\multirow{3}{*}{Hybrid} & Zamba2 & 1.2B, 2.7B & -- \\
 & Hymba & 1.5B & -- \\
 & Falcon-H1 & 0.5B, 1.5B & -- \\ \bottomrule
\end{tabular}%
}
\end{table}

\subsection{Evaluated Models}
\label{subsec:models}


%
To compare different architectural paradigms, we select a diverse set of representative, open-weight models.
Our selection spans three primary categories: Transformers, SSM, and Transformer-SSM hybrid model.
%
All models are evaluated using 16-bit bfloat (BF16) precision at first to reflect typical modern inference deployments.
A summary of the evaluated models is presented in
~\autoref{tab:model_zoo_short}.

\betterparagraph{Transformer Models} we select models from the Qwen2.5~\cite{yang2025qwen3} family, Phi-3-mini~\cite{abdin2024phi}, and Llama-3.2~\cite{grattafiori2024llama}, as they represent highly optimized, decoder-only architectures capable of processing long sequences.
We also include earlier models such as GPT2~\cite{radford2019language}, GPT-Neo~\cite{gpt-neo}, and TinyLlama~\cite{zhang2024tinyllama}, to provide insights on the limitation of traditional models (only supporting 1-2K sequence lengths on target platforms) . 
%
%

\betterparagraph{SSM Models} We select models from Mamba~\cite{gu2024mambalineartimesequencemodeling} and the newer Mamba-2~\cite{dao2024transformers} families with custom CUDA kernels for key operations like \texttt{causal depthwise conv1d} and the \texttt{parallel selective scan}. 
We analyze the performance of the official implementation with such custom CUDA kernels.
\betterparagraph{Hybrid Models} We evaluate three recent hybrid model families: \textbf{Zamba2}\cite{gloriosoZamba2SuiteTechnical}, \textbf{Hymba}\cite{dongHymbaHybridheadArchitecture2024} and \textbf{Falcon-H1}\cite{zuo2025falcon}.

\subsection{Model Selection Criteria}
\label{subsec:model_criteria}

The selection of these models was guided by several key principles to ensure a fair and relevant analysis for offline deployments.
First, we select models with comparable parameter counts across all three architectural categories to directly compare their efficiency.
Second, we include commonly adopted optimizations in our Transformer baselines, such as GQA~\cite{ainslie-etal-2023-gqa} and FlashAttention~\cite{dao2022flashattention}, to construct strong and competitive baseline against the cache-free SSMs.
Third, we prioritize models with officially available quantized checkpoints (e.g., INT4/INT8), as quantization is crucial for deploying LLMs on resource-constrained edge devices.
However, to ensure a fair baseline comparison, we report the comparative performance numbers for off-the-shelf models in default/BF16 precision in the following sections.
It is worth noting that Quantization yields consistent improvements in memory efficiency and supported context length across all evaluated architectures.
For example, comparing Mamba2-780m (BF16) to its quantized variant, Quamba2-780m (W4A8)\cite{chiang2025quamba2}, reveals a 3.5$\times$ reduction in model weight size (1488 vs. 424 MB).
At a prefill sequence length of 65K on an RTX 4090, the quantized model achieves a 1.26$\times$ speedup in TTFT (759 ms vs. 957 ms) and a 1.5$\times$ speedup in TPOT (1.52 ms vs. 2.28 ms).
%
Finally, we select SSMs and hybrid models that are capable of processing long input sequences without inherent architectural limitations.
%
We acknowledge that the open-source hybrid model space is still nascent, often lacking official quantized versions.
However, their inclusion is vital for understanding this promising architectural direction.

\subsection{Key Performance Metrics}
\label{subsec:metrics}



Our performance evaluation framework is designed to generate a rich set of statistics, capturing not only high-level performance but also fine-grained details essential for architectural comparison.
The output metrics and data from our framework are organized into three primary categories: computational performance (latency and throughput), memory, and energy metrics.

\betterparagraph{Computational Performance Metrics}
We measure end-to-end inference performance across different stages of generation and report Time to First Token (TTFT), Time per Output Token (TPOT), and throughput.
TTFT refers to the latency to process the input prompt (of varying length) and generate the initial output token, which corresponds to the latency of the compute-intense \textit{prefill} phase.
%
%
TPOT refers to the average latency to generate each subsequent token during the memory-intense \textit{decode} phase.
We compute decode throughput as $1 / \text{TPOT}$, whose unit is tokens per second.
To understand the performance characteristic across operators, we generate operator-wise latency breakdown for latency metrics.
%


\betterparagraph{Memory Metrics} As the memory capacity is a primary constraint on edge devices and a challenge for long context, we conduct  a thorough memory analysis, focusing on the peak memory usage.
To identify the peak memory usage, we record the maximum GPU memory reserved at the system level during inference, which includes model weights, activations (including KV cache, if any), and any framework overhead.
We also record operator-level peak memory usage to understand the operators' memory characteristics.
%


%
%

\betterparagraph{Energy Consumption} On edge deployment scenario, energy is an important metric since devices are often battery-operated. Therefore, we report the energy consumption, which we calculate from the power draw statistics over time (Energy = $\Sigma$ avg. power consumption in a time window $\times$ window size) using nvidia-smi tool.

\subsection{Workload Characterization flow}
\label{subsec:flow}

\insertFigure{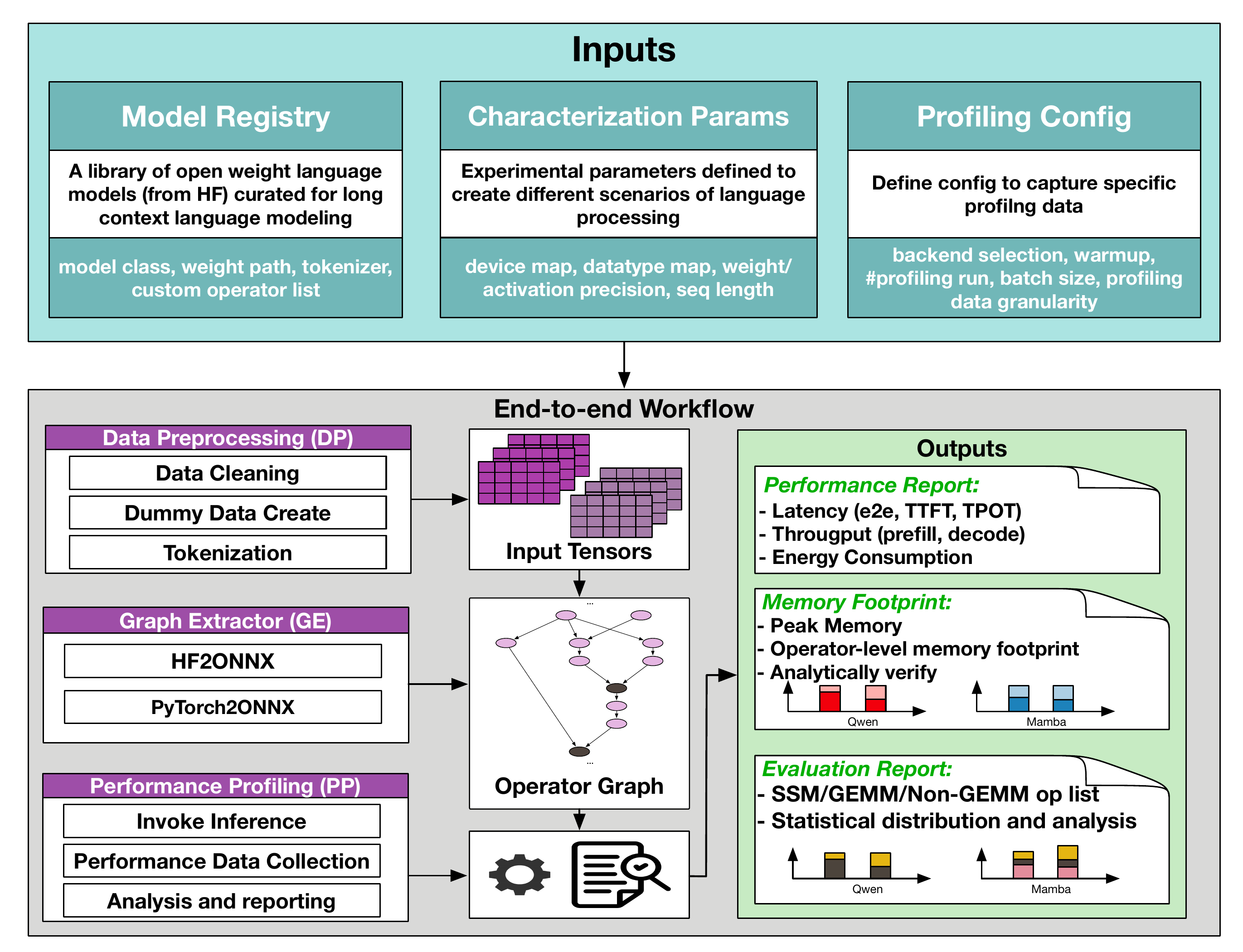}{Overall flow of the characterization framework for language models}

%
%
%

We illustrate our characterization framework in~\autoref{fig:flow_2a}.
Our flow takes workload configuration (model information in model registry and inference parameters such as sequence length), and profiling configurations.
The framework conducts necessary preprocessing (e.g., tokenization and graph extraction) and governs the performance profiling.
The output report includes end-to-end and operator level breakdown of the metrics we discussed in~\autoref{subsec:metrics}.
Note that the framework is highly configurable to support the rapidly evolving landscape of hybrid and traditional language models.
We discuss details of the each component in the input and the flow within our profiling framework.

\betterparagraph{Model Registry}The framework features a flexible model registry that can readily integrate new open-weight models from sources like the Hugging Face Hub.
Models compatible with HF Transformers package\cite{wolf2020transformers} are supported off-the-shelf including latest hybrids (Nemotron-H\cite{nvidiaNemotronHFamilyAccurate2025}, Falcon-H1R\cite{chaabane2026falcon}).
A new model is added by specifying its class, a link to its weights, and the corresponding tokenizer or other preprocessing steps.
Custom operators can be defined to accurately capture and classify their performance contribution.

\betterparagraph{Characterization Parameters}
Users can define various experimental parameters for comprehensive characterization.
These include the device map, data type mapping on the device (e.g. BF16, INT8, INT4), batch size and input/output sequence lengths to simulate various prefill and decode scenarios.
Across all characterization experiments, the batch size is strictly set to 1.
This configuration is chosen to accurately model single-user, interactive, on-device long-context inference scenarios.
To stress-test language models, our experiments are heavily focused on the prefill stage.
This design choice is motivated by the fact that reasoning-trace-based multi-turn conversations drastically increase the input context size over time.
Furthermore, when we experimented with balanced prefill-decode (e.g., 1024 input, 1024 output) and decode-heavy (e.g., 1024 input, 8192 output) scenarios by scaling the output sequence length ($S_{out}$), we observed that for models like Qwen2.5-0.5B and Mamba2-780m, the TPOT remains completely flat at 7.6 ms and 10.4 ms, respectively.
Because scaling the prefill sequence presents the primary bottleneck, our prefill-only experiments (e.g., presenting TTFT in \autoref{fig:intro_ttft_4}(a,c) or memory footprint in~\autoref{fig:memory_footprint_6}) use a single output token ($S_{out}=1$).
For generation scenarios presenting TPOT or throughput (such as \autoref{fig:energy_throughput}(b)), a fixed $S_{out}$ of 256 tokens is used, unless otherwise specified.
%
%
%
%
%
Similar evaluation strategies have been adopted in InfiniteBench\cite{zhang-etal-2024-bench}, LongBenchV2\cite{bai2025longbench} tasks (summarization, Q\&A) and MiniKV\cite{sharma-etal-2025-minikv} while scaling long-context inference.

\betterparagraph{Inference Backends and Profiling Configuration}
%
The framework is designed to be backend-agnostic.
It can be extended to support various inference frameworks like vLLM, TensorRT-LLM, or ONNX-Runtime to validate performance across different deployment toolchains and leverage backend-specific, model-agnostic runtime improvements.
For example, Qwen2.5-0.5B with 32K input length achieves a TTFT of 719ms in vLLM compared to 1250ms in PyTorch runtime.
%
%
Our experiments utilize the native PyTorch runtime not only for fairness or wide support but also to characterize intrinsic model architectural properties (operator bottleneck, memory-growth) rather than extrinsic system optimizations.
Efficient inference-engines like vLLM or TensorRT-LLM often fuse operations into monolithic kernels.
Furthermore, before starting the engine, vLLM reserves a large memory pool (typically ~90\% of the free GPU capacity) regardless of the active sequence length, which obfuscates fine-grained memory profiling.
The framework allows specifying warm-up and profiling iterations for robust and fair evaluation, with performance metrics averaged across runs.
Profiling data granularity can also be adjusted to capture performance breakdowns at various abstraction levels.

\betterparagraph{Characterization Workflow}
The framework's internal workflow is composed of different submodules.
The \textit{Graph Extractor (GE)} module extracts the computational graphs of input models using the graph exporter functionalities available in Hugging Face Transformers and PyTorch.
\textit{Data Preprocessing (DP)} module contains model-specific functions to fetch and clean raw data from real datasets (e.g., Wikitext) or create synthetic data (LongBenchV2\cite{bai2025longbench}, BABILong\cite{NEURIPS2024_babilong}-like), applying necessary transformations like tokenization to prepare the input tensors for the language models.
\textit{The Performance Profiling (PP)} module executes the inference runs, collects performance statistics, and generates output reports.
It selects the appropriate profiler based on the deployment flow. Operator-level latency is captured using the backend's native profiler (e.g., PyTorch Profiler, TensorRT's trtexec).
Peak memory footprint is tracked using PyTorch's CUDA-specific memory management functions, allowing us to monitor usage during distinct phases of inference.
%


Using the workflow, we conduct a thorough performance analysis of models listed in \autoref{tab:model_zoo_short}, on the consumer and edge platforms listed in \autoref{tab:gpu_setup}, which we discuss next.

\section{Case Studies}
\label{sec:case_studies}

\insertWideFigure{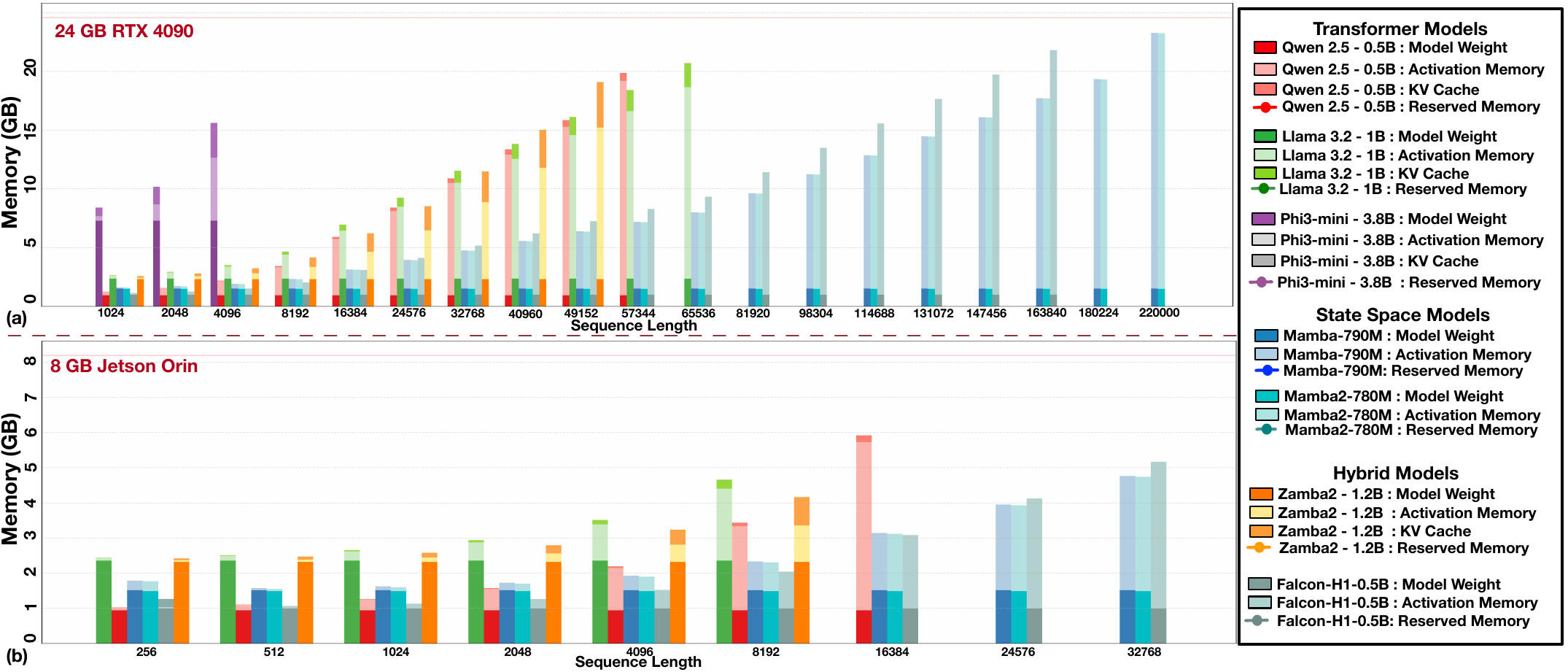}{
Memory footprint of prefill stage for Transformer, SSM, and Hybrid models on (a) consumer GPU (RTX 4090) and (b) edge GPU (Jetson Nano Orin)
}

%
We present performance analysis results with some highlighted cases that deliver new insights.
In \autoref{subsec:scalability_mem_footprint}, we investigate the memory footprint of the evaluated models and identify the maximum sequence length supported by each model on each evaluated platforms listed in~\autoref{tab:gpu_setup}.
We quantify the shift in energy consumption and end-to-end performance of all models in \autoref{subsec:energy_results}.
In \autoref{subsec:op_breakdown}, we perform an extensive performance characterization in the  operator granularity. 
Finally, we explore the performance implications of deploying these models on resource-constrained edge devices and provide a comparative analysis of the results.

\subsection{Memory Footprint and Scaling Limits}
\label{subsec:scalability_mem_footprint}

We present memory footprint characterization results over increasing sequence length in~\autoref{fig:memory_footprint_6}.
%
%
Our experiments focus on a single, long prefill stage to isolate the bottleneck in long-context applications (e.g., document summarization using RAG).
At the end of the prefill stage, memory pressure peaks because storing all prefill token activations ($S_{in}$) alongside the initial KV cache is required. 
In contrast, the decode stage processes tokens autoregressively, keeping only a single token’s activation in memory while the KV cache grows.
Reaching an OOM state during decode would require an excessively large number of output tokens ($S_{out}$).
%
%
%
%
Hence we scale the $S_{in}$ logarithmically from 1024 to 8192, then linearly in increments of 8192 up to 65536, and then in increments of 16384 until 180224.
We include additional data points at the end to capture behavior in some extreme cases.
We provide breakdowns into weight, activation, and KV cache (if applicable).
%
%
We stop plotting the data when we encounter the out-of-memory (OOM) issue.
For instance, the Phi-3-mini executes successfully at 4096 tokens but fails at 8192.
This indicates that its OOM threshold lies strictly between these two points.
%
%
%

From the Transformer category, we evaluated Qwen2.5 (0.5B), Llama3.2 (1B), and Phi-3-mini (3.82B).
While Phi-3 uses a classical decoder architecture, Qwen2.5 and Llama3.2 employ techniques like GQA and SwiGLU, which enhance scaling during inference.
On the consumer GPU (~\autoref{fig:memory_footprint_6}(a)), due to Phi-3's larger weight and KV cache size, Phi-3 was bottlenecked at a sequence length of 4096.
In contrast, Qwen and Llama extended much further, reaching 57,344 and 65,536 tokens, respectively.
For the SSMs, we considered Mamba and Mamba-2 (both \textasciitilde790M).
Although SSMs initially require higher memory footprint than Qwen, that of SSMs scales better with the lower activation memory growth and the absence of the KV cache.
%
Both Mamba and Mamba-2 are capable to handle up to 220K sequence length within the 24GB limit, which is unmatched by any similarly sized Transformer model.
This suggests that for contexts beyond \textasciitilde73K tokens, SSMs are the only viable architecture on consumer-grade GPUs without offloading.
We also benchmarked an earlier hybrid model Zamba2 (1.2B), which operated up to 49K tokens but showed significant KV cache consumption from its Transformer components due to not using GQA nor similar KV cache compression techniques.
%
A recently released hybrid model Falcon-H1 bridges this gap and can perform inference until \textasciitilde164K, way beyond Transformer models can reach with matching accuracy performance.
%
%

%
%

\betterparagraph{Impact of Efficient inference-engines}To compare these intrinsic limits in highly optimized production environments, we integrated these models into our framework with vLLM (chunked prefill enabled).
The OOM frontier for the Transformer-based Qwen extends significantly, going beyond the \textasciitilde57K token limit observed in native PyTorch to ~81K tokens in vLLM.
However, the maximum sequence lengths for Mamba and Falcon-H1 remain unchanged for the vLLM-support being experimental.

\betterparagraph{Impact of Quantization} 
In this case study, quantized checkpoints were not used because the available pretrained weights of different model classes use diverse and often incompatible quantization schemes.
To show the impact of quantization on memory scaling, we evaluated Qwen2.5-0.5B using GPTQ, a one-shot weight-only quantization.
This approach reduced the weight size by 3.6$\times$ (from 942 MB to 260 MB) while preserving activation memory (including the KV cache) in BF16 to maintain accuracy.
Despite the weight compression, the maximum supported sequence length for Qwen only increased slightly (from \textasciitilde57k to \textasciitilde59k).
For SSMs, we evaluated the model using Quamba2\cite{chiang2025quamba2} weights (W4A8).
At a 57K sequence length, Quamba2 saves 1.65$\times$ (4324 MB down to 2613 MB) on the total memory footprint.
This reduction significantly extends its OOM frontier.
While quantized checkpoints for the specific hybrid models evaluated are currently unavailable, we expect similar scaling trends when equivalent activation and weight quantization techniques are applied across architectures.

For the edge GPU (~\autoref{fig:memory_footprint_6} (b)), we exclude Phi-3 model as even its weights alone could not be accommodated within the available memory.
The primary observations from this experiment align with those from the consumer GPU, except for the significantly reduced operating range for all models.
On this severely constrained platform, SSMs and hybrids once again demonstrated a clear advantage, proving capable of processing contexts greater than 16K tokens while the Transformers could not.
\textbf{Overall, SSMs can operate at a maximum sequence length that is up to 4$\times$ greater than what is achievable by the Transformer models.} 

\subsection{Energy Consumption \& End-to-end Performance}
\label{subsec:energy_results}

\insertFigure{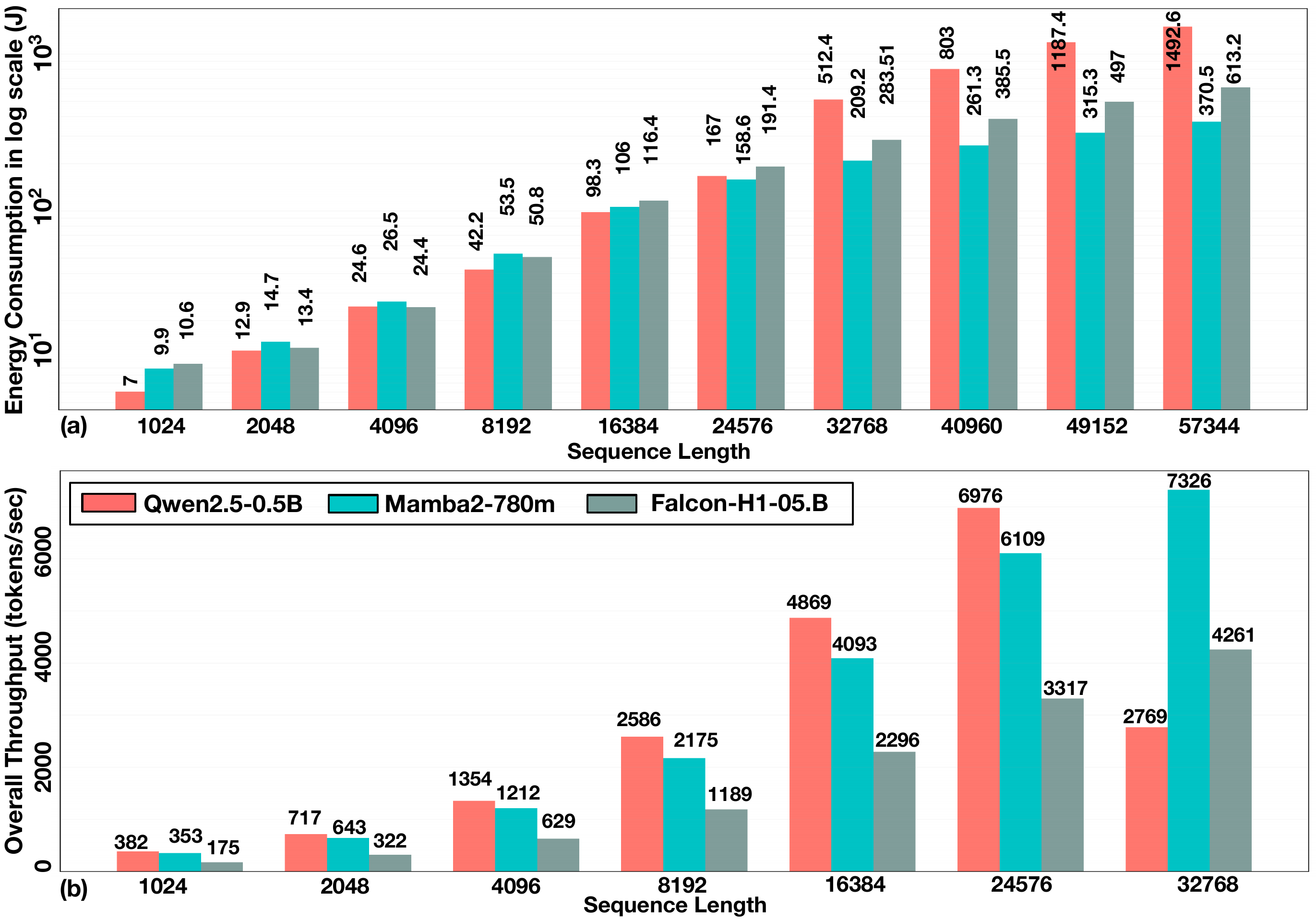}{(a) Inference energy consumption and (b) end-to-end throughput performance of Qwen2.5-0.5B (Transformer), Mamba2-780m (SSM), and Falcon-H1-0.5B (Hybrid) over sequence length on RTX4090}

\autoref{fig:energy_throughput}(a) presents the total energy consumption (in J) of the three models (Transformer, SSM, and Hybrid) when increasing sequence lengths from 1K to 57K.
Note that the Y-axis is presented in log scale to accommodate the wide range of the data.
In short contexts (<16K tokens), the Transformer model (Qwen2.5) consumes the least energy, benefiting from highly optimized GEMM kernels.
However, this trend sharply reverses as sequence length increases.
At 57K tokens, the quadratic computational burden of the Transformer results in high energy usage (1492 J), which is approximately \textbf{4$\times$} higher than the Mamba-2 SSM (370 J).
\textbf{The Hybrid Falcon-H1 model (613 J) effectively bridges this gap, offering a balanced profile that is 2.4$\times$ more energy-efficient than the Transformer} while retaining the accuracy gains discussed in \autoref{sec:background}.
%
Hybrid models are promising options for efficient and effective long-context inference, given their higher accuracy compared to SSMs.

We measure the overall throughput of text generation scenario (includes prefill and decode) across three model categories for different sequence lengths.
In our experiment, we consider batch sizes of 1 and 256 tokens as outputs.
Qwen outperforms in end-to-end performance with higher throughput in shorter sequences.
At 32K input sequence length when generating same output tokens, the TPOT jumps 2.4$\times$ more than the one in lower input sequence.
For SSM and Hybrids, TPOT remains fairly flat across experimented sequence lengths.
In \autoref{fig:energy_throughput}(b) \textbf{Mamba2 and Falcon-H1 attains 2.64$\times$ and 1.54$\times$ of the throughput of our baseline Transformer model at 32K sequence length}.

\subsection{Operator level performance}
\label{subsec:op_breakdown}
%
%
We investigate the performance implications of SSM-specific operations, such as \texttt{selective scan} and \texttt{causalconv1D}, under various deployment configurations, including sequence length changes and deployment platforms.
These novel operators cannot be straightforwardly classified as GEMM or non-GEMM.
At a finer granularity, they are composed of a mix of both types of operations.
For instance, low-level fused kernels like \texttt{mambainnerfn} or \texttt{mambasplitconv1dscancombinedfn} in popular SSM implementations often contain a diverse set of memory handling, element-wise arithmetic, and linear transformation operations.
Since these fused SSM operators contribute to the majority of the model's running time, it is more practical to analyze them as a separate, distinct category.
%
Some common activation functions like \textit{SiLU} are not be explicitly visible in the subsequent performance breakdown graphs due to its minor contribution to the total latency and/or operator fusion.
In this study, we focus on SSM and hybrid models as the bottlenecks of Transformers are well-studied~\cite{karami2025nongemmbenchunderstandingperformance}.

\insertFigure{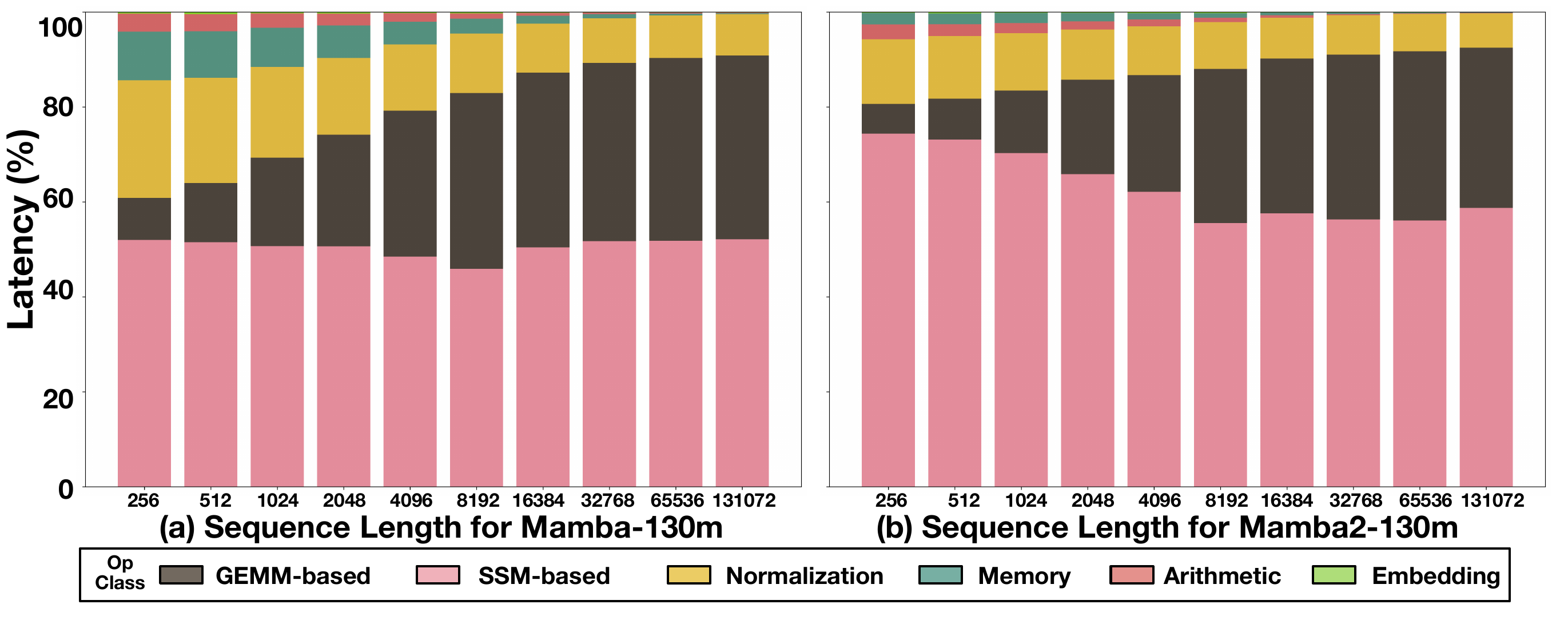}{Latency breakdown of SSMs into operator classes targeting RTX GPU on (a) the first generation of 130m mamba model and (b) second generation of the same family of model}

\subsubsection{SSMs on Consumer GPU}
\label{subsubsec:ssm_on_consumer}
To further understand model-specific characteristics, we analyze the operator-level latency breakdown of the state-of-the-art Mamba and Mamba-2 models as sequence length increases on the consumer-GPU, with results presented in \autoref{fig:mamba_breakdown_2a_singlecol}.
%
%
The input sequence length is increased logarithmically from 256 to 131072 tokens.
Each bar in the figure illustrates the percentage of total latency contributed by different operator categories.
The \textit{SSM-specific operators} are at the bottom, followed by \textit{GEMM-based operators}, and finally the \textit{non-GEMM operators}, which are sorted from highest to lowest contribution.

For the Mamba model, we observe a clear trend: as the sequence length grows, the relative latency of non-GEMM operators—such as normalization, memory, and arithmetic operations—steadily decreases.
Conversely, the contribution from GEMM operations (e.g., \texttt{matmul}, \texttt{linear}) monotonically increases, indicating that the workload becomes more compute-bound at longer sequences.
However, the latency share of the core SSM operators remains relatively flat.
%

When examining the newer Mamba-2 model under the same conditions, we observe similar trends.
However, the core SSM operator in Mamba-2 (\texttt{mambasplitconv1dscancombinedfn}) accounts for a larger portion of the total latency compared to its Mamba-1 counterpart (\texttt{mambainnerfn}).
This is due to architectural advancements in Mamba-2, which supports much larger state dimensions (from 16 to 64 or higher), leading to a more computationally intensive selective scan operation.
Furthermore, a key distinction emerges within the non-GEMM category: for Mamba-1 memory operators (transpose, slice, reshape etc.) are the dominant amongst non-GEMM component after normalization.
In contrast for Mamba-2, arithmetic operators consistently contribute more to the latency than memory ops.
This also can be attributed to Mamba-2's introduction of increased state-dimension and the multi-head structure, which expands the head dimension (e.g., from 1 to 64) compared to the earlier architecture.
%
%
%
%
%
%
%

Overall, we find that the performance of SSMs are dominated by newer SSM-specific operators, unlike Transformers that are usually dominated by GEMM or non-GEMM counterparts.
\textbf{This implies that optimizing SSM-specific operators is crucial for SSM performance for long context.}

\insertFigure{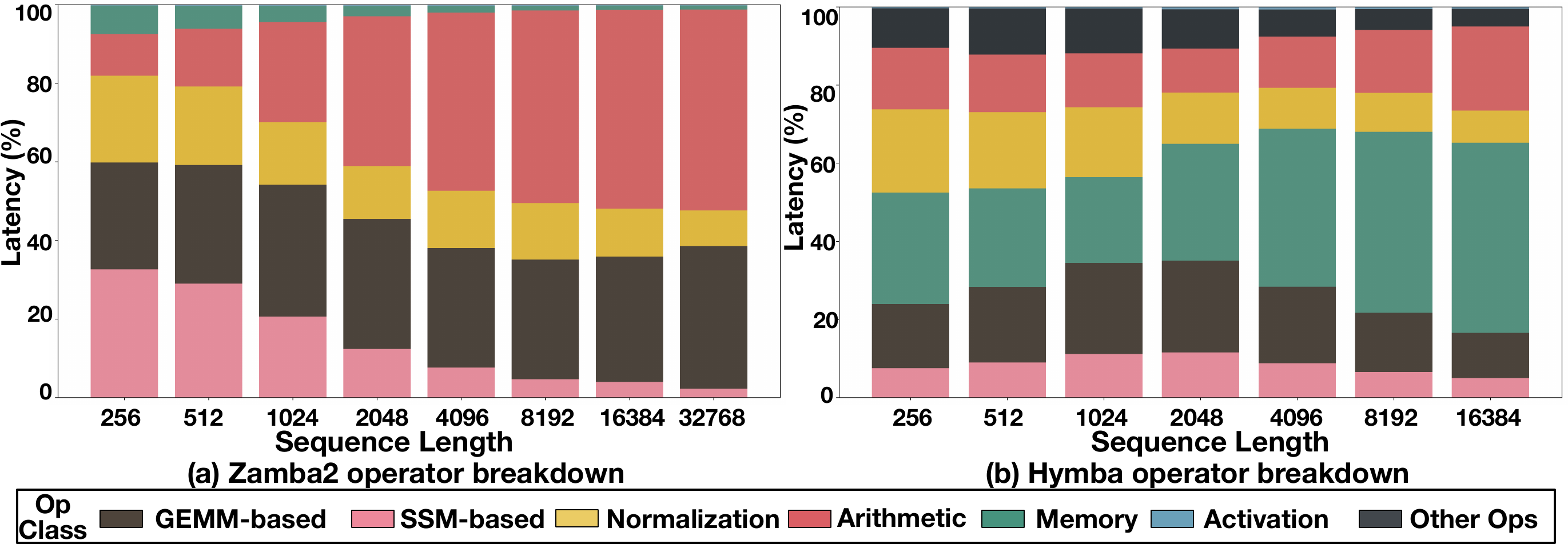}{Operator-wise latency breakdown for Hybrid models on consumer GPU targeting (a) Zamba2(1.2B params) and (b) Hymba (1.5B params).}
\subsubsection{Hybrid Models on Consumer GPU}
\label{subsubsec:hybrid_on_consumer}
We evaluate the latency breakdown into operators in hybrid models: Zamba2 (1.2B) and Hymba (1.5B)
, and present the results in \autoref{fig:hybrid_breakdown_2a_singlecol}.
The experimental setup remains consistent with our previous tests.
The input sequence length was doubled in each run until the consumer GPU ran out of memory.
%
Profiling these models often uncovers \texttt{cudaStreamSynchronize} calls (not unique to the hybrid models) present in the execution graph.
In hybrid models, these synchronizations are often artifacts of sub-optimal implementations.
For instance, in Zamba2, functions like \texttt{torch.is\_nonzero} return boolean values that require an explicit memory copy from the GPU to the CPU, which inherently triggers the synchronization bottleneck.
These synchronization points force the CPU to wait for GPU execution to complete, which can artificially inflate the reported latencies of certain operators.
%
%
%
The performance data we report combines both CPU and GPU time for each operator to provide a holistic view.

Our analysis reveals unique performance profiles for each hybrid model.
Zamba2 is built with Mamba2 backbone, so it is interesting to observe that the order of contribution of different operators in sub-categories is similar.
It uses full attention at shared layers without GQA or cross-layer cache sharing, resulting in notable KV cache pressure (leading to OOMs earlier, as seen in \autoref{fig:memory_footprint_6}(a)).
In contrast, Falcon-H1 utilizes a Mamba2 backbone with optimized attention that reduces KV cache, enabling longer sequences.
Hymba employs cross-layer KV cache sharing between consecutive layers and GQA within each layer.
This introduces additional memory copy and reference operations, which become visible as a pronounced bottleneck at high sequence lengths in \autoref{fig:hybrid_breakdown_2a_singlecol}(b).
It reveals an optimization potential in long-context scenarios for Hymba.
Furthermore, hybrid models often introduce context-switching overheads on the Streaming Multiprocessors (SMs), as the scheduler must alternate between compute-dense GEMM operations and memory-dense SSM warps, potentially disrupting pipeline efficiency.

%
%
%
%
%
\textbf{Unlike SSMs, Hybrid models' operation breakdown is not dominated by SSM-operators, but show variability related to rest of its architecture.}
This indicates that Hybrid models need careful characterization of performance to identify model-specific bottleneck, which our framework can assist.

\insertWideFigure{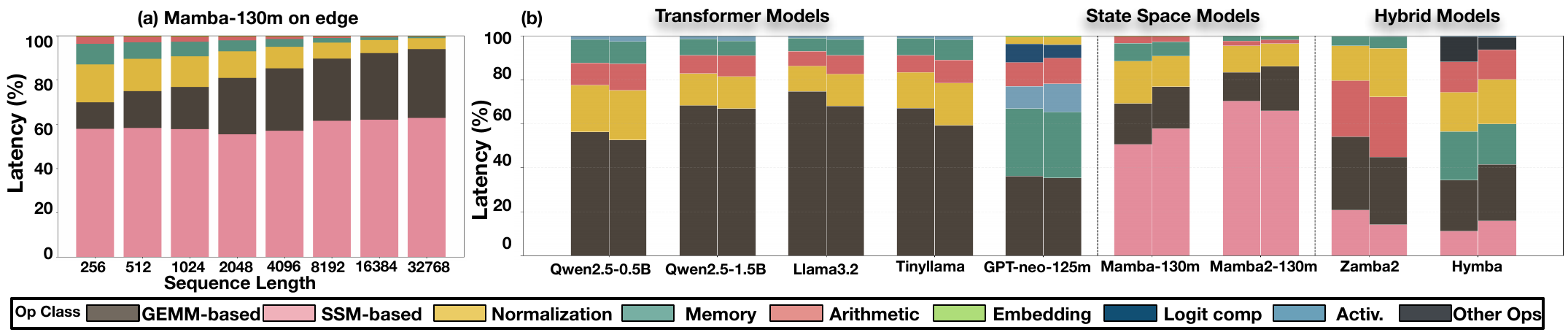}{(a) Latency breakdown results for Mamba-130m model in an edge GPU (Jetson Nano Orin), (b) Operator classes runtime breakdown comparison in two target platforms: Consumer GPU (left) and edge-GPU (right) for sequence length 1024}


\subsubsection{SSM on Edge GPU}
\label{subsec:ssm_on_edge}
Edge platforms present numerous constraints, including limited memory size and bandwidth and weak compute capability.
To investigate impact of such limitations, we deploy Mamba on a Nvidia Jetson Orin Nano listed in~\autoref{tab:gpu_setup}, pushing the input sequence length to its maximum supported limit while maintaining the same deployment scenario used for the consumer GPU.
We present the results in \autoref{fig:device_comparison_6}(a).
While the achievable sequence length on the edge GPU is smaller than that on the consumer GPU, the overall trends in operator-wise latency contribution are broadly similar, as shown in \autoref{fig:mamba_breakdown_2a_singlecol}.
The core SSM operator's relative contribution remains nearly flat.
One unique observation is that the SSM operators' latency share for mamba is consistently higher on the Jetson platform, accounting for over 55\% of the total runtime across all tested sequence lengths, which \textbf{imply the increased importance of optimizing SSM operators in some SSMs on edge platforms.}

However, we also observe that the significance of SSM operators change over model generations. 
For Mamba-2, SSM accounts for smaller latency share on the edge GPU than that on the consumer GPU.
However, the trend is opposite for the original Mamba.
Despite this shifting internal balance, the combined contribution of SSM and GEMM operators remains the dominant portion of the workload on both devices, consistently accounting for over \textbf{75-80\%} of the runtime.
This indicates that \textbf{significant optimization opportunities still exist within the core SSM operators} of such vanilla SSM architectures.

\subsubsection{Hybrid Models on Edge GPU}
\label{subsubsec:hybrid_on_edge}

As ~\autoref{fig:device_comparison_6} (b) presents that arithmetic and normalization operators become slightly more dominant while SSM operators become less significant compared to the consumer GPU case for Zamba 2. 
Hymba also presents overall similar trends as Zamba 2, while the significance of SSM operators increase, unlike Zamba 2.
Like the hybrid model on consumer edge GPU case, the results indicate that the performance bottleneck is model specific, and careful performance characterization is required for hybrids.

\subsubsection{Impact of Deployment Platforms}

We explore the performance impact from the choice of target platforms.
We deploy each evaluated model on both the consumer and the edge platforms with a fixed sequence length and present the detailed operator breakdown in \autoref{fig:device_comparison_6}(b).
For Transformers, the edge device requires more absolute time to complete individual GEMM operations due to its lower peak throughput.
%
However, when viewed holistically, the percentage of total runtime attributed to GEMM operations consistently declines across all Transformer-based models when deployed on the edge.
The counter-intuitive result is due to the \textbf{performance penalty for non-GEMM operators is more severe on the edge platform, causing their relative contribution to increase significantly}.
The specific non-GEMM bottleneck varies by model; for example, the dominant non-GEMM operator in Qwen is \textit{Normalization}, whereas for GPT-Neo, it is \textit{Memory}-related operations.
This underscores that optimizing non-GEMM operations is still a critical challenge for deploying Transformer models on edge devices.

For SSMs, as we discussed in~\autoref{subsubsec:ssm_on_consumer} and ~\autoref{subsec:ssm_on_edge},  we observe considerably different breakdown across two generations of the Mamba model family across platforms, while keeping SSM operators as the most dominant part.
We observe the similar trend across evaluated platforms, which indicate that \textbf{SSM operator is the prime optimization target regardless of the target platform}.

For hybrid models, as we discussed in~\autoref{subsubsec:hybrid_on_consumer} and~\autoref{subsubsec:hybrid_on_edge}, the performance profiles of recent hybrid models on different devices are  heavily model dependent, which requires target model-aware optimization.
We do not observe considerably different trend across deployment platforms, which implies that \textbf{optimization on one platform would benefit the performance on the other platform}.

\section{Related Works}
\label{sec:related_works}

\subsection{Performance Profiling of Language Models at the Edge}
\label{subsec:edge_profiling}
The performance characterization of deep learning workloads has been a central theme in computer architecture research, initially focusing on datacenter-scale training and inference\cite{qi2023performance}\cite{park2018deep}.
%
%
As running models on personal devices becomes more viable, research has shifted to profiling LLMs on edge and consumer hardware.
\cite{na2024understanding} have provided deep characterizations of LLM inference on modern CPUs with specialized matrix units, exploring alternatives to power-hungry GPUs.
Studies like \cite{10217850} have focused on mobile platforms, analyzing Transformer-based models on smartphone SoCs and the critical interplay between on-chip accelerators and memory subsystems.
%
%
Our work complements these efforts by shifting focus to emerging architectures and providing a comprehensive, comparative analysis of their memory and latency profiles during long-context inference in consumer devices.
%

\subsection{Characterization of Emerging Operators}
\label{subsec:emerging_operators}
To overcome the quadratic complexity of standard self-attention\cite{vaswani2017attention} and facilitate long-context inference, researchers developed more efficient variants.
These include sparse attention patterns, as seen in models like Longformer\cite{Beltagy2020Longformer}, and I/O-aware mechanisms like FlashAttention\cite{10.5555/3600270.3601459}, which reordered computations to minimize slow HBM memory accesses on GPUs.
The introduction of SSMs brought another class of specialized primitives, most notably the selective scan mechanism detailed in Mamba\cite{gu2024mambalineartimesequencemodeling}.
This operation combines parallel scans and convolutions, and its efficient implementation relies on custom-fused CUDA kernels optimized for modern GPUs\cite{mambarepo}.
These emerging operators represent a new frontier beyond the well-understood GEMM/non-GEMM dichotomy\cite{karami2025nongemmbenchunderstandingperformance}.
%
Our paper provides the first detailed, operator-level performance breakdown of the novel kernels in SSMs, specifically under the stress of long-sequence inputs.


\section{Conclusion and Future Work}
\label{sec:conclusion}


In this work, we presented a comprehensive performance and memory characterization of recent Transformer, State Space Model (SSM), and Hybrid architectures, targeting the growing need for long-context inference on consumer and edge GPUs.
%
%
Our findings confirm that SSMs without KV-cache and hybrids outperform Transformers for long contexts across platforms.
%
%
Furthermore, our deep operator-level characterization identified that custom kernels for SSM-specific operations are dominant in latency, especially on edge platforms.
Unlike SSMs, Hybrid models are more model-specific, which requires thorough performance profiling using a tool like ours.
We made our characterization tool open-source and believe it would be useful to guide optimizations of Hybrid models.



%
%
%


\bibliographystyle{IEEEtranS}
\bibliography{ref}
\appendices
\section*{Artifact Appendix}
\subsection{Abstract}
The \href{https://github.com/sapmitra/SSM-Scope}{artifact} contains the source code, pre-collected profiling data, and automated scripts necessary to reproduce the experimental results presented in this paper. It provides an end-to-end workflow to evaluate the performance, memory footprint, and energy consumption of State Space Models (SSMs), hybrid models, and Transformer-based models during long-context inference. The evaluation targets both consumer-grade (NVIDIA RTX 4090) and embedded (NVIDIA Jetson Nano Orin) hardware platforms.

The provided profiling scripts categorize the experimental results into three key areas: (1) \textbf{Inference Performance (\autoref{fig:intro_ttft_4}, \autoref{fig:accuracy_ttft_2},\autoref{fig:energy_throughput})}, which captures the long-context TTFT/TPOT performance reversal, accuracy-latency tradeoffs, and overall energy-throughput efficiency; (2) \textbf{Memory Footprint (\autoref{fig:memory_footprint_6})}, comparing Transformers’ out-of-memory boundaries to SSMs’ linear scaling; and (3) \textbf{Operator Breakdown (\autoref{fig:mamba_breakdown_2a_singlecol}, \autoref{fig:hybrid_breakdown_2a_singlecol}, \autoref{fig:device_comparison_6})}, detailing the execution latency bottlenecks of SSM and Hybrid models across both desktop and embedded GPU platforms.

\subsection{Artifact check-list (meta-information)}

{\small
\begin{itemize}
  \item {\textbf{Algorithm}: Transformer-based prefill/decode attention, SSM scans (Mamba, Mamba-2), and hybrid attention-SSM inference. }
  \item {\textbf {Program}: Python 3.10+, , CUDA 12.x (12.4 for current setup), PyTorch, vLLM (optional), Jupyter Notebook, shell script}
  \item {\textbf{Model:} Please refer to~\autoref{tab:model_zoo_short}}
  \item {\textbf {Data set:} For benchmarking synthetic random input tokens generated, like LongBenchV2\cite{bai2025longbench}.}
  \item {\textbf {Run-time environment:} Workstation: Linux Mint 21.1, Jetson Orin Nano: Ubuntu 22.04.5 (Jetpack 6.2).}
  \item {\textbf {Hardware:} Workstation: Intel i9-13900K, 64 GB DDR5 RAM, and Nvidia RTX 4090 24GB (PCIe)., Embedded GPU: Nvidia Jetson Nano Orin  }
  \item {\textbf {Execution:} Automated Scripts. Please refer to the main README file in the Github repository for getting started. Refer to the \href{https://github.com/sapmitra/SSM-Scope/blob/main/ispass_ae/scripts/env_setup/README.md}{env\_setup} first, then follow \href{https://github.com/sapmitra/SSM-Scope?tab=readme-ov-file#-paper-figures-at-a-glance}{README} of individual figures.}
  \item {\textbf {Metrics:} Time to First Token (TTFT), Time per Output Token (TPOT), Generation Throughput, Memory Footprint, and Power/Energy consumption.}
  \item {\textbf {Output:} Notebooks for easy visualization, CSV profile logs, JSON memory traces, and PNG figures matching those in the paper. }
  \item {\textbf {Experiments:} Please refer to \autoref{sec:intro}, \autoref{sec:background} and \autoref{sec:case_studies} for more details.  }
  \item {\textbf {How much disk space required (approximately)?:} Approximately 55 GB to store the models (each device), around 20 GB for python envs and the collected profiling traces.}
  \item {\textbf {How much time is needed to prepare workflow (approximately)?:} Approximately, setting up the workflow requires around 1--2 hours (primarily for downloading model weights and preparing venvs). }
  \item {\textbf {How much time is needed to complete experiments (approximately)?:} 10+ hours for full hardware profiling; $<20$ minutes to generate plots from the provided pre-collected data.}
  \item {\textbf {Publicly available?:} Yes (\href{https://github.com/sapmitra/SSM-Scope}{GitHub} and \href{}{Zenodo})}
  \item {\textbf {Code licenses (if publicly available)?:}  MIT License. }
  \item {\textbf {Archived (provide DOI)?:} 10.5281/zenodo.18917814}
\end{itemize}
}

\subsection{Description}

\subsubsection{How to access}
The source code is available on Zenodo at \href{https://doi.org/10.5281/zenodo.18917814}{https://doi.org/10.5281/zenodo.18917814}, and publicly hosted on Github at \href{https://github.com/sapmitra/SSM-Scope}{https://github.com/sapmitra/SSM-Scope}. 
Clone the repository using the following command:
\begin{verbatim}
git clone https://github.com/sapmitra/SSM-Scope.git
cd SSM-Scope
\end{verbatim}

\subsubsection{Hardware dependencies}
To reproduce the paper's results, the following systems are required: 
\begin{itemize}
\item{Edge System: Nvidia Jetson Orin Nano 8 GB. }
\item{Consumer System: Nvidia RTX 4090 24 GB GPU.}
\end{itemize}
Nevertheless, our workflow runs on any typical laptop, workstation, or server system with a CUDA-capable GPU. 

\subsubsection{Software dependencies}
\begin{itemize}
    \item Linux Mint 21.1 or Ubuntu 20.04 or 22.04 LTS.
    \item NVIDIA GPU Drivers and CUDA Toolkit ($\ge$ 12.1).
    \item Jetpack $\ge$6.2 for Jetson Nano Orin
    \item \href{https://github.com/state-spaces/mamba/releases/tag/v2.2.4}{mamba\_ssm} and \href{https://github.com/Dao-AILab/causal-conv1d/releases/tag/v1.5.0.post8}{causal-conv1d}
    \item \texttt{vLLM} (optional).
\end{itemize}

\subsubsection{Models Evaluated}
The evaluation scripts automatically download required models from Hugging Face via the \texttt{transformers} library. The evaluated models include:
Transformers (Llama-3.2-1B, Qwen2.5-1.5B, TinyLlama), SSMs (Mamba-130M, Mamba2-130M), and Hybrid Models (Zamba2, Hymba). Please refer to~\autoref{tab:model_zoo_short} for the detailed list.



\subsection{Installation}
Separate Python virtual environments are required because different LLMs have unique dependencies. Mamba~\cite{mambarepo} needs custom CUDA kernels, while Transformer-based models only need the HuggingFace stack. Detailed installation instructions are provided in \href{https://github.com/sapmitra/SSM-Scope/blob/main/ispass_ae/scripts/env_setup/README.md}{\texttt{ispass\_ae/scripts/env\_setup/}}.

\subsubsection{Transformers}
\small{
\begin{verbatim}
> python3 -m venv ~/.venvs/torch_transformers_ispass
> ## Activate the environment. 
> source ~/.venvs/torch_transformers_ispass/bin/activate
> pip install --upgrade pip
> ## Please check the README for rest of the exact
> ## required packages. 
\end{verbatim}
}

\subsubsection{Mamba Models}
\small{
\begin{verbatim}
> python3 -m venv ~/.venvs/torch_ssm_ispass
> source ~/.venvs/torch_ssm_ispass/bin/activate
> pip install --upgrade pip
> ## Please check the README for rest of the exact
> ## required packages.  
\end{verbatim}
}

\subsubsection{Falcon-H1/Hybrid Models}
\small{
\begin{verbatim}
> python3 -m venv ~/.venvs/torch_falcon_ispass
> source ~/.venvs/torch_falcon_ispass/bin/activate
> pip install --upgrade pip
> ## Please check the README for rest of the exact
> ## required packages. 
\end{verbatim}
}
\subsection{Experiment workflow}

The experimental workflow is designed to be automated and highly flexible. Evaluators can choose to either (a) regenerate all paper plots instantly using the pre-collected raw data, or (b) re-run the time-intensive hardware profiling pipeline from scratch.

\subsubsection{Generating Figures from Pre-collected Data}
Interactive data analysis is available via the provided Jupyter Notebooks in the \href{https://github.com/sapmitra/SSM-Scope/tree/main/ispass_ae/notebooks}{\texttt{ispass\_ae/notebooks/}} directory.
To facilitate rapid artifact evaluation, raw profiling results from our hardware runs are cached in the \href{https://github.com/sapmitra/SSM-Scope/tree/main/profile_data}{\texttt{profile\_data/}} (for RTX 4090) and \href{https://github.com/sapmitra/SSM-Scope/tree/main/profile_data_jetson}{\texttt{profile\_data\_jetson/}} (for Jetson) directories at the root of the repository.

\subsubsection{Executing Hardware Profiling from Scratch}
If a full reproduction on local hardware is desired navigate to the \href{https://github.com/sapmitra/SSM-Scope/tree/main/ispass_ae/scripts}{\texttt{ispass\_ae/scripts}} directory:
\begin{verbatim}
> cd <repo_root>ispass_ae/scripts/paper_figures/
\end{verbatim}
Within this directory, individual subfolders correspond to each key result figure from the paper (e.g., \texttt{Fig\_1}, \texttt{Fig\_3}, \texttt{Fig\_5a}). Each folder contains an automated bash script (e.g., \texttt{gen\_fig1.sh}) to reproduce the figures.
\small{
\begin{verbatim}
> cd <repo_root>/ispass_ae/scripts/paper_figures/Fig_1/
> chmod +x gen_fig1.sh
> bash gen_fig1.sh
\end{verbatim}
}
If \texttt{gen\_figx.sh} fails, please follow the README of each \texttt{Fig\_x} like \href{https://github.com/sapmitra/SSM-Scope/blob/main/ispass_ae/scripts/paper_figures/Fig_1/README.md}{this} one to follow data generation and plotting scripts.

\subsection{Evaluation and expected results}

Running the scripts in every figure subdirectory will reproduce~\autoref{fig:intro_ttft_4}, ~\autoref{fig:accuracy_ttft_2}, ~\autoref{fig:memory_footprint_6}, ~\autoref{fig:energy_throughput}, ~\autoref{fig:mamba_breakdown_2a_singlecol}, ~\autoref{fig:hybrid_breakdown_2a_singlecol} and ~\autoref{fig:device_comparison_6}. 

The scripts will generate corresponding data in (\texttt{src/tpot\_logs, src/profiling\_logs, src/energy\_logs, src/throughput\_logs etc}) and the plots in the respective Figure directories (\texttt{<repo\_root>/ispass\_ae/scripts/paper\_figures/Fig\_x}).
The reproduced latency results are expected to be close to the results in the paper, but not an exact match because of potential differences in the hardware or software environment. 
%

\subsection{Methodology}

Submission, reviewing and badging methodology:

\begin{itemize}
  \item \url{https://www.acm.org/publications/policies/artifact-review-and-badging-current}
  \item \url{https://cTuning.org/ae}
\end{itemize}

\end{document}